\documentclass{article}
\usepackage{amssymb}
\usepackage[dvips]{graphicx}
\usepackage{amsmath}
\usepackage{graphicx}
\usepackage{makeidx}

\begin{document}
\title{Solitons in combined linear and nonlinear lattice potentials}
\author{Hidetsugu Sakaguchi$^{1}$ and Boris A. Malomed$^{2}$\\
$^{1}$Department of Applied Science for Electronics and Materials,\\
Interdisciplinary Graduate School of Engineering Sciences,\\ Kyushu
University, Kasuga, Fukuoka 816-8580, Japan\\
$^{2}$Department of Physical Electronics,\\ School of Electrical Engineering,
Faculty of Engineering, \\Tel Aviv University, Tel Aviv 69978, Israel}
\maketitle
\small
\bigskip
abstract\\
We study ordinary solitons and gap solitons (GSs) in the framework of the
one-dimensional Gross-Pitaevskii equation (GPE) with a combination of linear
and nonlinear lattice potentials. The main points of the analysis are
effects of (in)commensurability between the lattices, development of
analytical methods, \textit{viz}., the variational approximation (VA) for
narrow ordinary solitons, and various forms of the averaging method for
broad solitons of both types, and also the study of mobility of the
solitons. Under the direct commensurability (equal periods of the lattices, $%
L_{\mathrm{lin}}=L_{\mathrm{nonlin}}$), the family of ordinary solitons is
similar to its counterpart in the GPE without external potentials. In the
case of the subharmonic commensurability, with $L_{\mathrm{lin}}=(1/2)L_{%
\mathrm{nonlin}}$, or incommensurability, there is an existence
threshold for the ordinary solitons, and the scaling relation
between their amplitude and width is different from that in the
absence of the potentials. GS families demonstrate a bistability,
unless the direct commensurability takes place. Specific scaling
relations are found for them too. Ordinary solitons can be readily
set in motion by kicking. GSs are mobile too, featuring inelastic
collisions. The analytical approximations are shown to be quite
accurate, predicting correct scaling relations for the soliton
families in different cases. The stability of the ordinary solitons
is fully determined by the VK (Vakhitov-Kolokolov) criterion, i.e.,
a negative slope in the dependence between the solitons's chemical
potential $\mu $ and norm $N$. The stability of GS families obeys an
inverted (``anti-VK") criterion, $d\mu /dN>0$, which is explained by
the approximation based on the averaging method. The present system
provides for a unique possibility to check the anti-VK criterion, as
$\mu (N)$ dependences for GSs feature turning points, except for the
case of the direct commensurability.

\section{Introduction}

It is well known that periodic potentials, induced by optical lattices
(OLs), provide for a versatile tool for controlling dynamics of
Bose-Einstein condensates (BECs). This tool is especially efficient for the
creation and stabilization of solitons -- both ordinary ones and gap
solitons (GSs), which are supported by the interplay of the OL potential and
self-repulsive nonlinearity. Many results obtained in theoretical and
experimental studies of this topic were summarized in reviews focused on
one-dimensional (1D) \cite{reviews-1D} and multidimensional \cite%
{review-multiD} matter-wave dynamics. Earlier, a similar model was
introduced in optics for the description of spatial solitons in nonlinear
waveguides with a periodic transverse modulation of the refractive index
\cite{Wang}. In the experiment, lattices controlling the transmission of
optical beams were implemented in the form of photoinduced gratings in
photorefractive crystals, which made it possible to create various species
of 1D and 2D spatial solitons \cite{photorefr}. Gratings were also created
as permanent structures written by femtosecond laser beams in silica \cite%
{Jena}. Recently, this topic was reviewed in Ref. \cite{Barcelona-review}; a
closely related topic is the study of discrete solitons in optics, which was
a subject of another comprehensive review in Ref. \cite{discrete-review}.

A different possibility, which has drawn much attention in studies of BEC
(thus far, primarily at the theoretical level), is the use of a spatially
profiled effective nonlinearity, that may be implemented by means of
properly designed configurations of external fields via the
Feshbach-resonance effect. In terms of the condensed-matter theory, the
nonuniform nonlinearity coefficient induces an effective \textit{%
pseudopotential} \cite{pseudo,Dong}, that can be used to control the
dynamics of localized modes. Various problems of this sort were considered
in 1D settings, with periodic pseudopotentials in the form of nonlinear
lattices (NLs) \cite{we1,NL-1D,Kwok}, as well as with spatial modulations of
the nonlinearity coefficient represented by one \cite{1delta} or two \cite%
{Dong} delta-functions (actually, the model with the self-attractive
nonlinearity concentrated at a single delta-function was introduced long ago
as a model for tunneling of interacting particles through a junction \cite%
{Azbel}). In particular, the configuration with two delta-functions makes it
possible to study spontaneous symmetry breaking of matter waves trapped by a
symmetric double-well pseudopotential \cite{Dong}. Specially chosen profiles
of the nonlinearity coefficient may also be employed to design a
pulse-generating atomic-wave laser \cite{laser}, as well as traps and
barriers for such pulses \cite{trap}. The analysis of 1D matter-wave
solitons in NLs was further extended for two-component models \cite%
{NL-1D-2comp,vector}, and for some spatiotemporal patterns of the
nonlinearity modulation \cite{spatiotemp}. Certain results for solitons
supported by 2D nonlinear pseudopotentials were reported too, although the
stabilization of 2D solitons in this setting is a tricky problem \cite{NL-2D}%
.

In addition to the BEC, a periodic modulation of the nonlinearity is
possible in optics, where it was analyzed in terms of temporal \cite%
{temporal} and spatial \cite{NL-optics,Kominis} solitons. A
discussion of practical possibilities to create NLs in optical media
in the ``pure form" (without affecting the linear properties, i.e.,
the refractive index) can be found in recent paper \cite{vector}. An
experimental observation of NL-supported optical solitons (in the
form of
surface solitons at an interface between lattices) was reported in Ref. \cite%
{experiment}.

A natural generalization of the study of NL pseudopotentials is to consider
atomic and optical media equipped with a combination of nonlinear and linear
lattices, in 1D \cite{Panos,Spain,Kwok} and 2D \cite{NL-OL2D} settings. In
the experiment, this may be implemented by applying the above-mentioned
techniques simultaneously -- for instance, combining the OL, which induces
the linear periodic potential in the BEC, and the patterned magnetic field,
which gives rise to the nonlinear pseudopotential via the Feshbach
resonance. Actually, photonic-crystal fibers, where various species of
spatial solitons have been predicted \cite{PhotCryst}, also belong to this
type of media. Under special conditions, the combined models admit
analytical solutions, see Ref. \cite{Kwok} and references therein. In
particular, exact solutions were elaborated in detail for the lattices of
the Kronig-Penney (piecewise-constant) type \cite{Kominis}.

The objective of this work is to investigate the existence, stability, and
mobility of ordinary 1D solitons and their GS counterparts in the combined
NL-OL model, with emphasis on effects of the \emph{commensurability}
(spatial resonance) between the nonlinear and linear lattices. In the
general case, the solitons are found in a numerical form. For narrow
ordinary solitons (those whose chemical potential, $\mu $, falls into the
semi-infinite gap of the linearized version of the model), we also develop a
variational approximation (VA). For broad solitons, both ordinary ones and
GSs with $\mu $ belonging to the first finite bandgap, we elaborate
analytical approximations based on different versions of the averaging
technique.

The stability of the solitons is investigated below by means of systematic
simulations of their perturbed evolution. We conclude that the well-known
VK\ (Vakhitov-Kolokolov) criterion\ \cite{VK} completely determines the
actual stability of the ordinary solitons (the criterion states that a
necessary stability condition is a negative slope in the dependence of the
solitons' chemical potential, $\mu $, on their norm, $N$, i.e., $d\mu /dN<0$%
). For the GS families, our analysis leads to a different conclusion: their
stability fully obeys an ``anti-VK" criterion, \textit{viz}%
., $d\mu /dN>0$. As a matter of fact, all stable GS families in previously
studied models had only the positive slope of $\mu (N)$, thus satisfying the
latter criterion automatically. However, the present model offers a rather
unique chance to test it in a nontrivial situation, when $\mu (N)$ curves
for the GSs may have portions with both positive and negative slope,
separated by turning points. Using the averaging method, we also produce a
justification for the anti-VK criterion, which is relevant, at least, for
subfamilies of broad GSs.

The rest of the paper is organized as follows. The model is formulated in
the next section, which also reports basic numerical results. In the case of
the direct commensurability between the nonlinear and linear lattices (for
those with equal periods, $L_{\mathrm{lin}}=L_{\mathrm{nonlin}}$), it is
concluded that ordinary solitons are similar to soliton solutions of the
nonlinear Schr\"{o}dinger equation (NLSE) in the free space (without an
external potential), in the sense that there is no threshold for their
existence, the entire soliton family is stable, and the amplitude and width
of the soliton, $A$ and $W$, obey the usual scaling relation, $A\propto 1/W$%
. On the contrary to that, there is an existence threshold for ordinary
solitons in the case of the incommensurability, or if the commensurability
(spatial resonance) between the lattices is \textit{subharmonic}, with $L_{%
\mathrm{lin}}=L_{\mathrm{nonlin}}/2$. In those cases, the scaling relation
between the amplitude and width is different too -- featuring, in
particular, $A\propto 1/W^{2}$ for the subharmonic resonance. The same
commensurability-depending change in the scaling is observed in GS families,
which also demonstrate a bistability in cases different from the direct
commensurability.

Analytical results, which may explain a considerable part of the
numerical findings, are collected in Section III. First, we develop
the VA for narrow ordinary solitons, and then the averaging method
is reported, in different forms for different cases. It is
demonstrated that both the VA and averaging produce results which
are in good agreement with numerical observations, in relevant
regions of the parameter space. In particular, as briefly mentioned
above, the averaging lends an explanation to the ``anti-VK"
stability criterion for GSs. In the case of $L_{\mathrm{lin}}=L_{\mathrm{%
nonlin}}/2$, an average equation for the envelope amplitude includes a
quintic nonlinear term, rather than the cubic one, which accounts for the
above-mentioned change in the scaling relation between $A$ and $W$. The
paper is concluded by Section IV.

\section{Numerical results}

\subsection{The model}

The model combining the linear OL potential and nonlinear NL pseudopotential
is based on the known effectively one-dimensional Gross-Pitaevskii equation
for the mean-field wave function, $\phi \left( x,t\right) $ \cite%
{Kominis,Panos,Spain,Kwok}. In the scaled form, the equation is
\begin{equation}
i\phi _{t}=-\left( 1/2\right) \phi _{xx}-\left[ \epsilon \cos (2\pi x)+g\cos
(\pi qx)~|\phi |^{2}\right] \phi ,  \label{1}
\end{equation}%
where $\epsilon $ is the strength of the linear OL, and $g=\pm 1$ is fixed
by the normalization. The NL wavenumber is $q$, the above-mentioned periods
of the linear and nonlinear lattices being $L_{\mathrm{lin}}\equiv 1$ and $%
L_{\mathrm{nonlin}}=2/q$. For $q=0$ and $g=-1$, Eq. (\ref{1}) amounts to the
NLSE with the OL potential and constant coefficient in front of the
self-defocusing cubic term. As is well known, this equation supports
families of stable GS solutions \cite{reviews-1D,we2}. On the other hand, in
the absence of the OL, $\epsilon =0$, the NL\ supports ordinary solitons,
but not GSs \cite{we1,NL-1D,Kwok}.

Stationary solutions with chemical potential $\mu $ are looked for in the
form of $\phi (x,t)=u(x)\exp (-i\mu t)$, where $u(x)$ is a real function.
The substitution of this expression into Eq.~(\ref{1}) yields an ordinary
differential equation,
\begin{equation}
\mu u=-\left( 1/2\right) u^{\prime \prime }-\left[ \epsilon \cos (2\pi
x)+g\cos (\pi qx)u^{2}\right] u,  \label{2}
\end{equation}%
which can be derived from the corresponding Lagrangian,
\begin{equation}
2L=\int_{-\infty }^{+\infty }\left[ \mu u^{2}-\left( 1/2\right) \left(
u^{\prime }\right) ^{2}+\epsilon \cos \left( 2\pi x\right) ~u^{2}+\left(
g/2\right) \cos \left( \pi qx\right) ~u^{4}\right] dx.  \label{L}
\end{equation}%
We have constructed numerical solutions for localized stationary
modes by means of the shooting method applied to Eq. (\ref{2}). The
stability of the so found solutions against small perturbations was
tested through direct simulations of Eq.~(\ref{1}). The results were
also compared to predictions of the VK criterion for the ordinary
solitons, and to the above-mentioned ``anti-VK" criterion for GSs.
Numerical results reported below are obtained for the OL strength
$\epsilon =5$, which adequately represents the generic situation,
for the ordinary solitons and GSs alike.

\begin{figure}[tbp]
\begin{center}
\includegraphics[height=4.cm]{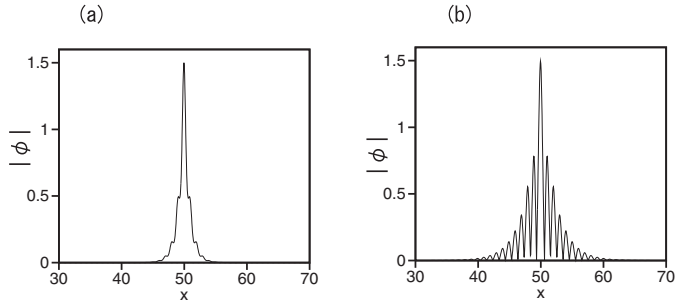}
\end{center}
\caption{Typical examples of a stable broad ordinary soliton (a) and stable
gap soliton (b), at $g=+1$ and $-1$, respectively, in the case of the
nonlinear lattice with period $L_{\mathrm{nonlin}}=2$ (i.e., $q=1$). The
amplitudes of both solitons are $A=1.5$. The chemical potential and norm
are, respectively, $\protect\mu =-1.198$ and $N=1.628$ for the ordinary
soliton, and $\protect\mu =2.629$ and $N=2.176$ for the gap soliton.}
\label{f1}
\end{figure}

Figures \ref{f1}(a) and (b) display, respectively, examples of ordinary
solitons (for $g=+1,~q=1$) and GSs (for $g=-1,~q=1$) with equal amplitudes.
The choice of $q=1$ corresponds to the above-mentioned case of the
subharmonic resonance between the OL and NL, $L_{\mathrm{lin}}=(1/2)L_{%
\mathrm{nonlin}}$. In this case, stable ordinary solitons and GSs coexist
for either sign of $g$. If, for instance, $g=+1$, ordinary solitons are
located around even sites of the NL, $x=2n$ (with integer $n$), while GSs
may be centered at odd sites, $x=2n+1$.

\subsection{Ordinary solitons}

\begin{figure}[tbp]
\begin{center}
\includegraphics[height=4.cm]{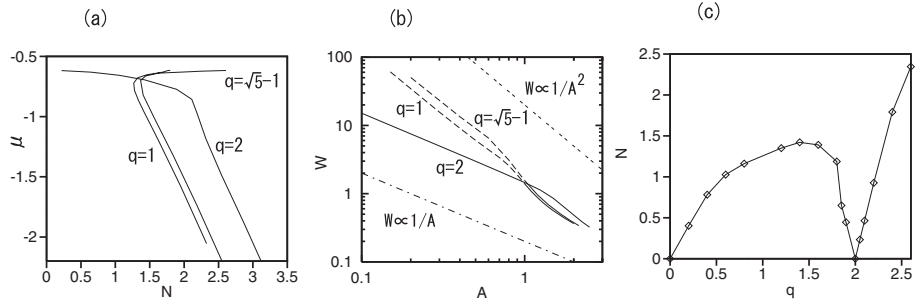}
\end{center}
\caption{(a) Chemical potential $\protect\mu $ versus norm $N$ for families
of ordinary solitons at different values of $q$ and $g=+1$. Portions of the
curves with $d\protect\mu /dN<0$ and $d\protect\mu /dN>0$ represent,
respectively, stable and unstable (sub)families, in agreement with the VK
criterion. (b) The log-log plot of amplitude $A$ versus width $W$ [the
latter is defined as per Eq. (\protect\ref{Width})]. Bold dashed lines
correspond to unstable branches of the soliton families, for $q=1$ and $%
\protect\sqrt{5}-1$. Thin dashed reference lines designate scalings which
different families obey at small values of $A$, namely, $W\propto 1/A$ and $%
W\propto 1/A^{2}$. (c) The stability boundary for the ordinary solitons,
which is defined, as per the VK criterion, by condition $dN/d\protect\mu =0$%
. As predicted by the criterion and verified in direct simulations, the
solitons are stable above the boundary.}
\label{f2}
\end{figure}

Figure \ref{f2}(a) represents a family of the ordinary solitons, by means of
the relation between their norm, $N=\int_{-\infty }^{+\infty }u^{2}(x)dx$,
and chemical potential $\mu $, at three characteristic values of the NL's
wavenumber, $q=1,\sqrt{5}-1$, and $2$, for $g=+1$. These values are chosen
because $q=2$ corresponds to the direct commensurability between the linear
and nonlinear lattices ($L_{\mathrm{lin}}=L_{\mathrm{nonlin}}$), $q=1$
represents, as said above, the subharmonic commensurability [$L_{\mathrm{lin}%
}=\left( 1/2\right) L_{\mathrm{nonlin}}$], and $q=\sqrt{5}-1$ corresponds to
incommensurate lattices. All values of $\mu $ for the ordinary solitons fall
into the semi-infinite gap of the spectrum induced by the OL potential in
the linearized version of Eq. (\ref{1}).

At $q=2$, direct simulations demonstrate that the entire soliton family is
stable, precisely as suggested by the VK criterion, $d\mu /dN<0$, see Fig. %
\ref{f2}(a). For $q=1$ and $q=\sqrt{5}-1$, the results are different,
featuring non-monotonous relations $\mu (N)$. Accordingly (again in the
agreement with the prediction of the VK criterion), narrow solitons with
larger amplitudes, which correspond to the branches of $\mu (N)$ with $d\mu
/dN<0$ in Fig. \ref{f2}(a), are stable, while their loosely bound (broad)
counterparts, corresponding to the branches with $d\mu /dN>0$, are unstable.
Another difference from the case of the direct commensurability is that
there is a minimum \ value of the norm, $N_{\min }$ (the threshold), which
is necessary for the existence of the ordinary solitons at $q=1$ and $\sqrt{5%
}-1$, while there is no threshold at $q=2$.

In fact, the situation in the case of $q=1$ and $\sqrt{5}-1$ -- the
existence of the threshold value, $N_{\min }$, which separates stable and
unstable branches of the ordinary-soliton solutions -- is qualitatively
similar to what is known in the model with the NL but no linear potential
\cite{we1}. On the other hand, the situation in the case of $q=2$ -- the
absence of $N_{\min }$ and the existence of the single branch of the soliton
solutions, which is entirely stable -- resembles well-known properties of
soliton solutions of the NLSE without any lattice, linear or nonlinear.

\begin{figure}[tbp]
\begin{center}
\includegraphics[height=4.cm]{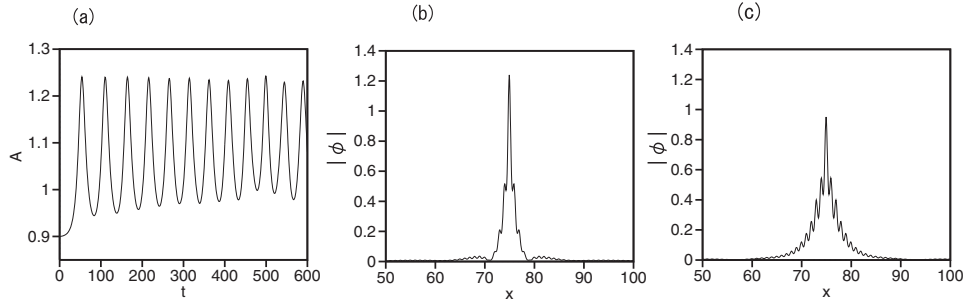}
\end{center}
\caption{An example of the spontaneous rearrangement of an unstable ordinary
soliton, with amplitude $A=0.9$ and norm $N=1.32,$ into a robust breather,
at $q=1$ and $g=+1$. (a) The evolution of the soliton's amplitude. (b) The
field profile, $\left\vert \protect\phi (x,t)\right\vert $, at $t=110$. (c)
The profile at $t=140$. Panels (b) and (c) display the shape of the breather
at points where its width is close, respectively, to the minimum and maximum
values.}
\label{f3}
\end{figure}

The evolution of those ordinary solitons which are unstable is illustrated
in Fig. \ref{f3} for $q=1$ and $g=+1$. It is observed that the unstable
soliton, with initial amplitude $A=0.9$, rearranges itself into a narrower
persistent breather, with time-average amplitude $A_{\mathrm{br}}\approx 1.1$%
, while the norm is kept constant. The transformation of the unstable
ordinary solitons into stable breathers is also similar to what was reported
in the model without the OL \cite{we1}.

Properties of the ordinary solitons in the same three families, with $q=1,%
\sqrt{5}-1$ and $2$ (and $g=+1$), are further illustrated in Fig.~\ref{f2}%
(b) through relations between the soliton's amplitude, $A$, and its width, $%
W $, which we define by
\begin{equation}
W^{2}=N^{-1}\int_{-\infty }^{+\infty }|\phi (x)|^{2}\left( x-L/2\right)
^{2}dx  \label{Width}
\end{equation}%
($x=L/2$ is the central point of the integration domain). At $q=2$, relation
$W(A)$ features scaling $W\propto 1/A$ for relatively small values of $A$.
This is the same scaling as featured by exact soliton solutions of the NLSE
in the free space, which is in line with the above observation that the
soliton family at $q=2$ is similar to that in the NLSE without any lattice.
However, the scaling is different in the other families, featuring $%
W(A)\propto 1/A^{2}$ for $q=1$, and $W(A)\propto 1/A^{1.8}$ for $q=\sqrt{5}%
-1 $. The two latter scaling relations imply that the width of the
respective solitons is essentially larger than in their free-space
counterparts, therefore we call them broad solitons.\

Figure \ref{f2}(c) summarizes the results by means of the stability boundary
for the ordinary solitons in the plane of $\left( q,N\right) $. The boundary
is identified as a VK-critical curve, along which $d\mu /dN$ vanishes, the
stability area (with $d\mu /dN<0$) being located above the curve. Systematic
simulations performed in regions below and above the boundary have confirmed
that the solitons in these regions are, respectively, unstable \ and stable
(unstable solitons transform themselves into breathers, as shown in Fig. \ref%
{f3}). The stability area reaches the limit of $N=0$ (very broad solitons
with a vanishingly small amplitude) in the form of the cusps in Fig. \ref{f2}%
(c) at $q=0$ and $q=2$. Recall that $q=0$ with $g=+1$ corresponds to the
constant coefficient of the self-attractive nonlinearity in Eq. (\ref{1}),
while $q=2$ corresponds to the direct commensurability between the linear
and nonlinear lattices.

\subsection{Static gap solitons}

\begin{figure}[tbp]
\begin{center}
\includegraphics[height=4.cm]{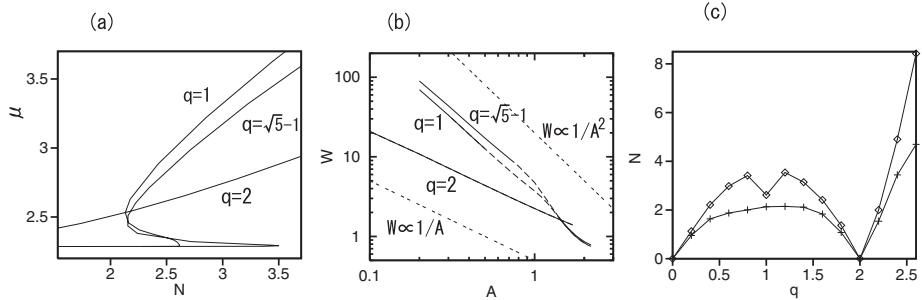}
\end{center}
\caption{(a) The relation between norm $N$ and chemical potential $\protect%
\mu $ of gap solitons for $q=1,\protect\sqrt{5}-1$, and $2$, for
$g=-1$. Portions of the curves with $d\protect\mu /dN>0$ and
$d\protect\mu /dN<0$ represent stable and unstable (sub)families,
respectively, obeying the ``anti-VK" criterion (see the text). (b)
The log-log plot of amplitude $A$ versus width $W$, for the gap
solitons. Bold dashed portions of the curves correspond to unstable
solutions, with $d\protect\mu /dN<0$. Thin dashed reference lines
designate scalings $W\propto 1/A$ and $W\propto 1/A^{2}$. (c)
Critical lines $dN/d\protect\mu =0$ in the plane of $\left(
q,N\right) $. There is a single stable gap soliton above the upper
line and beneath the lower one, and three solutions -- two stable
and one unstable -- in the bistability region between the two
lines.} \label{f4}
\end{figure}

In this work, the consideration of GS families was confined to the
first finite bandgap induced by the OL potential, in terms of the
linearized version of Eq. (\ref{1}). The $N(\mu )$ and $W(A)$ curves
for these families are displayed in Figs. \ref{f4}(a) and (b), for
the same three cases as above, \textit{viz}., $q=1,\sqrt{5}-1$ and
$2$, fixing $g=-1$ [width $W$ is again defined as per Eq.
(\ref{W})]. The VK criterion does not apply to GSs. Nevertheless,
the results strongly suggest that the stability of all GS families
follows an ``\textit{anti-VK}" condition, $d\mu /dN>0$. For
subfamilies of broad GSs, this condition is derived below from an
effective envelope equation for broad GSs which amounts to an
``inverted" NLSE, with the self-repulsive nonlinear term and
negative effective mass, see Eqs. (\ref{Phi-gap}). In that
approximation, GSs reduce to ordinary solitons if the wave function
is subjected to the complex conjugation, which implies the inversion
of the sign of $\mu $, and of $d\mu /dN$ as well.

In accordance with what is said above, the entire GS family for $q=2$, which
satisfies the ``anti-VK" condition everywhere in Fig. \ref%
{f4}(a), is found to be completely stable in direct simulations. On
the other hand, the $\mu (N)$ curves for $q=1$ and $q=\sqrt{5}-1$
feature two folds, and direct simulations corroborate the
instability of portions of the GS families with $d\mu /dN<0$. To the
best of our knowledge, the present model produces the first example
of $\mu (N)$ characteristics for GSs with turning points, which
makes the anti-VK criterion amenable to the actual verification. In
the standard model with the constant nonlinearity coefficient
\cite{Kivshar,reviews-1D}, as well as in its version with the
quasiperiodic OL potential \cite{we3}, the monotonous character of
the curves does not allow the verification of the criterion.

The stability diagram in the plane of $\left( q,N\right) $, as
predicted by the ``anti-VK" criterion, is displayed in Fig.
\ref{f4}(c).
It includes two critical curves with $dN/d\mu =0$. As suggested by Fig. \ref%
{f4}(b), and completely confirmed by systematic simulations, in the regions
above the top curve and below the bottom one there is a single solution,
which is stable. Between the curves, there are three solutions, two of which
are stable, i.e., this is a \textit{bistability} region. The critical curves
feature cusps near $q=0$ and $q=2$, similar to the situation displayed in
Fig.~\ref{f2}(c).

From Fig. 4(b) we conclude that the scaling relations between the GS's width
and amplitude for broad solitons (with small values of $A$) take the
following form: For $q=2$, $W\sim 1/A$; for $q=\sqrt{5}-1$, $W\propto
1/A^{1.85}$; and for $q=1$, $W\propto 1/A^{2}$, i.e., almost exactly the
same as their counterparts for the ordinary solitons, see above. All
portions of the GS families obeying these scaling relations are stable. On
the other hand, the simulations demonstrate that unstable GSs (those with $%
d\mu /dN<0$) are not transformed into breathers, unlike unstable ordinary
solitons, but rather suffer a gradual decay into quasi-linear waves (not
shown here).

\section{Analytical methods}

\subsection{The variational approximation for ordinary solitons}

Narrow stationary solitons of the ordinary type (corresponding to $g=+1$)
can be naturally approximated by means of the VA, using the simplest
Gaussian ansatz \cite{Wang},
\begin{equation}
u(x)=A\exp \left[ -x^{2}/\left( 2W^{2}\right) \right] ,  \label{ans}
\end{equation}%
with norm $N=\sqrt{\pi }A^{2}W.$ The substitution of ansatz (\ref{ans}) in
Lagrangian (\ref{L}) yields the effective Lagrangian, written in terms of
the norm, instead of amplitude $A$:
\begin{equation}
2L_{\mathrm{eff}}=\mu N-\frac{N}{4W^{2}}+\epsilon Ne^{-\pi ^{2}W^{2}}+\frac{%
N^{2}}{2\sqrt{2}W}e^{-\pi ^{2}q^{2}W^{2}/8}.  \label{Leff}
\end{equation}%
Variational equations following from Eq. (\ref{Leff}), $\partial L_{\mathrm{%
eff}}/\partial W=0$ and $\partial L_{\mathrm{eff}}/\partial N=0$, take the
form of
\begin{eqnarray}
4\pi ^{2}\epsilon W^{4}e^{-\pi ^{2}W^{2}}+\left( 1/\sqrt{2\pi }\right)
\left( 1+\pi ^{2}k^{2}W^{2}\right) NWe^{-\left( \pi qW\right) ^{2}/8} &=&1,
\label{W} \\
\left( 4W^{2}\right) ^{-1}-\epsilon e^{-\left( \pi W\right) ^{2}}-\left( 1/%
\sqrt{2\pi }\right) \left( N/W\right) e^{-\left( \pi qW\right) ^{2}/8}
&=&\mu .  \label{mu}
\end{eqnarray}

Figure 5 compares the $\mu (N)$ curves produced by Eqs. (\ref{W}) and
( \ref{mu}) to their numerically found counterparts, for the NL wavenumbers $%
q=1$ and $2$. It is seen that the VA provides for a good approximation for
narrow solitons with values of $\mu $ which are not too close to the edge of
the semi-infinite gap. Portions of the curves corresponding to broad
solitons, which are located near the gap's edge, are not captured by the VA,
as the actual shape of these solitons is different from the Gaussian, see,
e.g., Fig. \ref{f1}(a).
\begin{figure}[tbp]
\begin{center}
\includegraphics[height=4.cm]{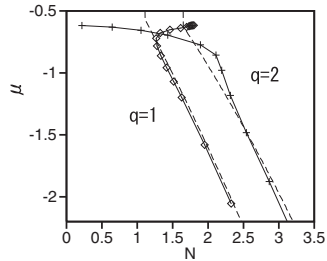}
\end{center}
\caption{Comparison of the variational (dashed lined) and numerically found
(chains of symbols) curves $\protect\mu (N)$ for ordinary solitons ($g=+1$).}
\label{f7}
\end{figure}

\subsection{The averaging method}

\subsubsection{Ordinary solitons}

The approximation based on averaging can be applied to broad solitons, which
have a small amplitude and large norm. In the case of the ordinary solitons (%
$g=+1$), this is the situation opposite to that (narrow localized modes) for
which the VA was presented in the previous section. To develop the averaging
approach for ordinary solitons, we adopt the ansatz
\begin{equation}
\phi (x,t)=\Phi \left( x,t\right) \left[ 1+2\alpha \cos (2\pi x)\right] ,
\label{envelope}
\end{equation}%
where the slowly varying amplitude function, $\Phi $, multiplies the
simplest approximation for the Bloch wave function which may be used near
the edge of the semi-infinite gap, with $\alpha =\left( \sqrt{\pi
^{4}+\epsilon ^{2}/2}-\pi ^{2}\right) /\epsilon $ (this approximation is
obtained by dint of the analysis presented in Ref. \cite{we2}). The
substitution of ansatz (\ref{envelope}) into Eq. (\ref{1}) and averaging,
also performed along the lines of Ref. \cite{we2}, lead to the asymptotic
NLSE for the slowly varying envelope function,
\begin{equation}
i\frac{\partial \Phi }{\partial t}=-\frac{1}{2m_{\mathrm{eff}}^{\mathrm{(ord)%
}}}\frac{\partial ^{2}\Phi }{\partial x^{2}}+g_{\mathrm{eff}}^{\mathrm{(ord)}%
}|\Phi |^{2}\Phi ,  \label{Phi}
\end{equation}%
where the calculations yield the following coefficients:
\begin{equation}
m_{\mathrm{eff}}^{\mathrm{(ord)}}=\frac{2\pi ^{4}+\epsilon ^{2}+\pi ^{2}%
\sqrt{4\pi ^{4}+2\epsilon ^{2}}}{10\pi ^{4}+\epsilon ^{2}-3\pi ^{2}\sqrt{%
4\pi ^{4}+2\epsilon ^{2}}},  \label{m-ord}
\end{equation}%
\begin{equation}
g_{\mathrm{eff}}^{\mathrm{(ord)}}=-\frac{\left\langle \left[ 1+2\alpha \cos
(2\pi x)\right] ^{4}\cos (q\pi x)\right\rangle }{1+2\alpha ^{2}},
\label{g-ord}
\end{equation}%
with $\left\langle ...\right\rangle $ standing for the spatial average.
Obviously, the effective nonlinearity coefficient given by expression (\ref%
{g-ord}) vanishes unless $q$ takes values $0,2,4,6,$ or $8$. In particular, $%
g_{\mathrm{eff}}^{\mathrm{(ord)}}=-(1+12\alpha ^{2}+6\alpha ^{4})/(1+2\alpha
^{2})$ for $q=0$, and $g_{\mathrm{eff}}^{\mathrm{(ord)}}=-4\alpha (1+3\alpha
^{2})/(1+2\alpha ^{2})$ for $q=2$. Actually, $g_{\mathrm{eff}}^{\mathrm{(ord)%
}}$ does not vanish at $q=2n$ with any integer $n$, if higher-order
harmonics are kept in the expansion of the Bloch function at the edge of the
semi-infinite gap, $1+2\sum_{n}\alpha _{n}\cos (2\pi nx)$, cf. the
lowest-order approximation used in Eq. (\ref{envelope}). With the value of
the OL strength adopted in the numerical simulations reported above, $%
\epsilon =5$, Eq. (\ref{m-ord}) yields $m_{\mathrm{eff}}^{\mathrm{(ord)}%
}\approx 1.128$, which is virtually identical to the numerically found
effective mass, see Eq. (\ref{meff2}).

The soliton solution to Eq. (\ref{Phi}) with an arbitrary amplitude, $A$, is
\begin{equation}
\Phi =A\exp \left( \frac{i}{2}g_{\mathrm{eff}}^{\mathrm{(ord)}}A^{2}t\right)
~\mathrm{sech}\left( \sqrt{g_{\mathrm{eff}}^{\mathrm{(ord)}}m_{\mathrm{eff}%
}^{\mathrm{(ord)}}}Ax\right) .  \label{sol-ord}
\end{equation}%
This solution explains scaling $W\propto 1/A$, which is observed in Fig.~\ref%
{f2}(b) for broad ordinary solitons in the case of $q=2$. Figure 6
displays a direct comparison of profiles of a typical broad ordinary
soliton, as obtained in the numerical form and produced by the averaging
method. Good agreement between the two profiles is obvious [note that the
figure displays the full wave function, $\left\vert \phi (x)\right\vert $,
rather than the envelope, $\Phi (x)$, see Eq. (\ref{envelope})].

\begin{figure}[tbp]
\begin{center}
\includegraphics[height=4.cm]{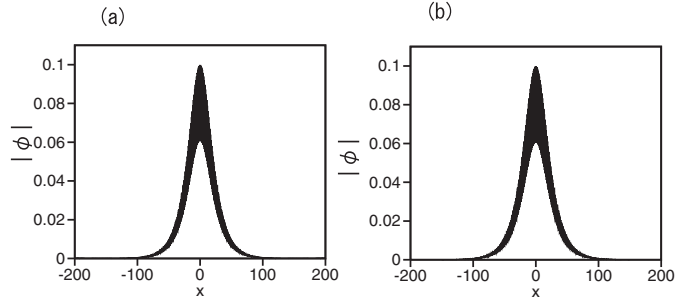}
\end{center}
\caption{(a) The numerical obtained profile of an ordinary soliton with
amplitude $A=0.1$ for $g=+1$ and $q=2$. (b) The profile predicted for the
same soliton\ by means of the averaging method, see Eqs. (\protect\ref%
{envelope}), (\protect\ref{sol-ord}), (\protect\ref{m-ord}), and (\protect
\ref{g-ord}).}
\label{f8}
\end{figure}

\subsection{Gap solitons}

\subsubsection{Direct commensurability between the nonlinear and linear
lattices}

To apply the averaging approximation to GSs, which correspond to $g=-1$ in
Eq. (\ref{1}), we follow Ref. \cite{we2} and adopt the simplest ansatz which
is relevant in this case,
\begin{equation}
\phi \left( x,t\right) =\Phi \left( x,t\right) \cos (\pi x),
\label{envelope-gap}
\end{equation}%
with a slowly varying amplitude $\Phi \left( x,t\right) $. The difference in
the form of the carrier wave in this expression, in comparison to Eq. (\ref%
{envelope}), is due to the fact that, near the edge of the first finite
bandgap, the Bloch function is close to a periodic function, whose period
is twice as large as that of the underlying OL potential. On the contrary,
near the edge of the semi-infinite gap the period of the Bloch function
coincides with the OL's period.

The substitution of ansatz (\ref{envelope-gap}) in Eq. (\ref{1}) and the
application of the averaging method yields the respective asymptotic NLSE
equation for the amplitude function,%
\begin{equation}
i\frac{\partial \Phi }{\partial t}=-\frac{1}{2m_{\mathrm{eff}}^{\mathrm{(gap)%
}}}\frac{\partial ^{2}\Phi }{\partial x^{2}}+g_{\mathrm{eff}}^{\mathrm{(gap)}%
}|\Phi |^{2}\Phi ,  \label{Phi-gap}
\end{equation}%
cf. Eq. (\ref{Phi}). The effective mass and interaction coefficients in Eq. (%
\ref{Phi-gap}) are found to be
\begin{equation}
m_{\mathrm{eff}}^{\mathrm{(gap)}}=\epsilon /\left( \epsilon -2\pi
^{2}\right) ,  \label{m-gap}
\end{equation}%
\begin{equation}
g_{\mathrm{eff}}^{\mathrm{(gap)}}=2\left\langle \cos (\pi x))^{4}\cos (q\pi
x)\right\rangle .  \label{g-gap}
\end{equation}%
Note that coefficient (\ref{g-gap}) is different from zero only for three
values of $q$, \textit{viz}., $g_{\mathrm{eff}}^{\mathrm{(gap)}%
}(q=0)=3/4,~g_{\mathrm{eff}}^{\mathrm{(gap)}}(q=2)=1/2,$ and $g_{\mathrm{eff}%
}^{\mathrm{(gap)}}(q=4)=1/8$. Together with
\begin{equation}
m_{\mathrm{eff}}^{\mathrm{(gap)}}\approx \allowbreak -0.339,
\label{meff-gap}
\end{equation}%
which Eq. (\ref{m-gap}) yields for $\epsilon =5$ (recall this value
of the OL strength is fixed in the present work), the \emph{complex
conjugate} form \ of Eq. (\ref{Phi-gap}) gives rise to the usual
soliton solutions for $\Phi ^{\ast }$, cf. solutions (\ref{sol-ord})
for $\Phi $. This fact gives the explanation to the scaling
$W\propto 1/A$ for the broad GSs, as observed in Fig. \ref{f4}(b)
for $q=2$.\ For stationary solutions, the complex conjugation
implies, as mentioned above, the reversal of the sign of the
chemical potential. This explains why the stability of the broad GSs
obeys the ``anti-VK" criterion, $d\mu /dN>0$, which is simply the
reverse of the ordinary negative-slope VK condition for the stable
solutions of the equation for $\Phi ^{\ast }$.

The above description is also relevant for moving GSs. In particular, the
comparison of the analytically predicted effective mass (\ref{meff-gap}) to
the empirical dynamical mass (\ref{meff}), drawn for the moving soliton from
numerical data at $q=2$, clearly demonstrates the high accuracy of the
averaging approximation for the broad GSs, both static and moving ones.

\subsubsection{Subharmonic commensurability between the nonlinear and linear
lattices}

The above approximation for GSs does not produce any definite result for $%
q=1 $, when the period of the NL in Eq. (\ref{1}) is twice as large as OL's
period (the subharmonic resonance between the nonlinear and linear lattices,
as defined above). To derive an effective envelope equation in this case, we
notice that the substitution of original ansatz (\ref{envelope-gap}) in Eq. (%
\ref{1}) and making use of the same effective mass as given by Eq. (\ref%
{m-gap}), but without averaging the nonlinear term, gives rise to the
following equation:
\begin{equation}
i\frac{\partial \Phi }{\partial t}=-\frac{1}{2m_{\mathrm{eff}}^{\mathrm{(gap)%
}}}\frac{\partial ^{2}\Phi }{\partial x^{2}}+\cos ^{3}(\pi x)~|\Phi
|^{2}\Phi ,
\end{equation}%
where it is taken into regard that $q=1$. This equation suggests that ansatz
(\ref{envelope-gap}) should be replaced by the following one, for stationary
solutions:
\begin{equation}
\phi \left( x,t\right) =e^{-i\mu t}\left[ \Phi _{1}\left( x\right) \cos (\pi
x)+\Phi _{4}\left( x\right) \cos ^{4}\left( \pi x\right) \right] ,
\label{14}
\end{equation}%
where both functions $\Phi _{1}$ and $\Phi _{4}$ are slowly varying ones.
The substitution of ansatz (\ref{14}) into Eq. (\ref{1}) and straightforward
trigonometric expansions make it possible to eliminate $\Phi _{4}$ in favor
of $\Phi _{1}$:%
\begin{equation}
\Phi _{4}(x)=\left( \mu -\frac{\pi ^{2}}{2m_{\mathrm{eff}}^{\mathrm{(gap)}}}%
\right) ^{-1}\left\vert \Phi _{1}(x)\right\vert ^{2}\Phi _{1}(x).
\end{equation}%
The remaining equation for $\Phi _{1}(x)$ is the NLSE with the \emph{quintic}
self-focusing nonlinear term:
\begin{equation}
\mu \Phi _{1}=-\frac{1}{2m_{\mathrm{eff}}^{\mathrm{(gap)}}}\frac{\partial
^{2}\Phi _{1}}{\partial x^{2}}-\frac{15m_{\mathrm{eff}}^{\mathrm{(gap)}}}{%
8\left( \pi ^{2}-2m_{\mathrm{eff}}^{\mathrm{(gap)}}\mu \right) }|\Phi
_{1}|^{4}\Phi _{1}.  \label{quintic}
\end{equation}
An obvious soliton solution to Eq. (\ref{quintic}) with arbitrary amplitude $%
A$ is
\begin{equation}
\Phi _{1}=A~\sqrt{\mathrm{sech}\left( kx\right) },~k=\sqrt{5}m_{\mathrm{eff}%
}^{\mathrm{(gap)}}\left( \pi ^{2}-2m_{\mathrm{eff}}^{\mathrm{(gap)}}\mu
\right) ^{-1/2}A^{2},~\mu =-\left( 8m_{\mathrm{eff}}^{\mathrm{(gap)}}\right)
^{-1}k^{2}.  \label{sol-q}
\end{equation}%
Solution (\ref{sol-q}) yields the scaling relation $W\propto 1/A^{2},$ which
explains the same scaling that was observed, at $q=1$, for broad GSs in Fig.~%
\ref{f4}(b).

It is necessary to mention that the nonstationary version of Eq. (\ref%
{quintic}), with $\mu \Phi _{1}$ replaced by $i\partial \Phi _{1}/\partial t$%
, corresponds to the one-dimensional NLSE with the critical
(quintic, in the 1D case) self-focusing nonlinearity, whose
solutions, tantamount to those given by Eq. (\ref{sol-q}), are the
so-called ``one-dimensional Townes solitons". It is well known that
the entire family of such solitons is unstable (see, e.g., Ref.
\cite{Salerno}). Nevertheless, simulations of Eq. (\ref{1})
demonstrate that the GSs approximated by the asymptotic solution
(\ref{sol-q}) form a \emph{stable} family, as long as
the solitons remain broad, see the corresponding stable branch in Fig. \ref%
{f4}(b). Thus, the asymptotic description of the broad GSs by means of Eq. (%
\ref{quintic}) is valid, at $q=1$, only for static solutions, while their
dynamical behavior does not obey the straightforward time-dependent version
of this equation.

\section{Conclusion}

We have investigated the existence, stability and mobility of ordinary
solitons and GSs (gap solitons) in the 1D model combining nonlinear and
linear periodic lattices in the Gross-Pitaevskii equation for the nearly
one-dimensional BEC. The emphasis was made on the study of effects of the
commensurability and incommensurability between the lattices, as well as on
the development of analytical methods -- the VA (variational approximation)
for narrow ordinary solitons, and averaging method for broad solitons of
both types. We have demonstrated that, in the case of the direct
commensurability between the lattices (equal periods), the ordinary solitons
are similar to their counterparts in the free space. In the case of the
subharmonic commensurability and incommensurability, the situation is
different, featuring the existence threshold for the solitons, and a
different scaling relation between their amplitude and width. Similar
scaling relations are found for GS families, which demonstrate the
bistability in cases different from the direct commensurability. Ordinary
solitons may travel long distances, if kicked. Broad GSs are mobile too,
although collisions between them are inelastic.

The analytical approximations demonstrate good accuracy in
appropriate parameter regions. In particular, they correctly explain
different scaling relations for the soliton families at different
commensurability orders. As concerns the stability of the ordinary
solitons, it is accurately predicted by the VK criterion.
Simultaneously, the stability of GSs obeys the ``anti-VK" criterion,
an explanation to which was given by means of the effective equation
produced through the averaging method. A notable feature of the
present model is that it gives rise to characteristics $\mu (N)$ for
the GSs that feature turning points (except for the case of the
direct commensurability). Unlike previously studied models with the
constant nonlinearity coefficient, the presence of the turning
points has made it possible to test the anti-VK stability criterion
for the GS families.

The analysis reported in this work may be naturally extended by
studying GSs in higher finite bandgaps; in particular, a challenging
issue is to verify the ``anti-VK" stability criterion in the higher
gaps. It may also be interesting to develop a systematic analysis of
the commensurability for solitons and solitary vortices in the 2D
model based on the combination of linear and nonlinear lattices.


\begin{thebibliography}{99}
\bibitem{reviews-1D} V. A. Brazhnyi and V. V. Konotop, Mod. Phys. Lett. B
\textbf{18}, 627 (2004); O. Morsch and M. Oberthaler, Rev. Mod. Phys.
\textbf{78}, 196 (2006).

\bibitem{review-multiD} B. A. Malomed, D. Mihalache, F. Wise, and L. Torner,
J. Optics B: Quant. Semicl. Opt. \textbf{7}, R53 (2005).

\bibitem{Wang} B. A. Malomed, Z. H. Wang, P. L. Chu, and G. D. Peng, J. Opt.
Soc. Am. B \textbf{16}, 1197 (1999).

\bibitem{photorefr} N. K. Efremidis, S. Sears, D. N. Christodoulides, J. W.
Fleischer, and M. Segev, Phys. Rev. E \textbf{66}, 046602 (2002); J. W.
Fleischer, T. Carmon, M. Segev, N. K. Efremidis, and D. N.
Christodoulides, Phys. Rev. Lett. \textbf{90}, 023902 (2003); J. W.
Fleischer, M. Segev, N. K. Efremidis, and D. N. Christodoulides, Nature
\textbf{422}, 147 (2003); D. Neshev, E. A. Ostrovskaya, Y. Kivshar, and W. Kr%
\'{o}likowski, Opt. Lett. \textbf{28}, 710 (2003); D. N. Neshev, T. J.
Alexander, E. A. Ostrovskaya, Y. S. Kivshar, H. Martin, I. Makasyuk, and Z.
Chen, Phys. Rev. Lett. \textbf{92}, 123903 (2004); J. W. Fleischer, G.
Bartal, O. Cohen, O. Manela, M. Segev, J. Hudock, and D. N. Christodoulides,
\textit{ibid}. \textbf{92}, 123904 (2004); Z. Chen, H. Martin, E. D.
Eugenieva, J. Xu, and A Bezryadina, \textit{ibid}. \textbf{92}, 143902
(2004); Z. Chen, M. Stepi\'{c}, C. R\"{u}ter, D. Runde, D. Kip, V.
Shandarov, O. Manela, and M. Segev, Opt. Express \textbf{13}, 4314 (2005);
R. Fischer, D. Tr\"{a}ger, D. N. Neshev, A. A. Sukhorukov, W. Kr\'{o}%
likowski, C. Denz, and Y. S. Kivshar, Phys. Rev. Lett. \textbf{96}, 023905
(2006).

\bibitem{Jena} A. Szameit, D. Bl\"{o}mer, J. Burghoff, T. Schreiber, T.
Pertsch, S. Nolte, and A. T\"{u}nnermann, Opt. Express \textbf{13}, 10552
(2005); D. Bl\"{o}mer, A. Szameit, F. Dreisow, T. Schreiber, S. Nolte, and
A. T\"{u}nnermann, \textit{ibid}. \textbf{14}, 2151 (2006); A. Szameit, J.
Burghoff, T. Pertsch, S. Nolte, A. T\"{u}nnermann, and F. Lederer, \textit{%
ibid} \textbf{14}, 6055 (2006).

\bibitem{Barcelona-review} Y. V. Kartashov, V. A. Vysloukh, and L. Torner,
Progress in Optics \textbf{52}, 63 (ed. by E. Wolf: North Holland,
Amsterdam, 2009).

\bibitem{discrete-review} F. Lederer, G. I. Stegeman, D. N. Christodoulides,
G. Assanto, M. Segev, and Y. Silberberg, Phys. Rep. \textbf{463}, 1 (2008).

\bibitem{pseudo} W. A. Harrison, \textit{Pseudopotentials in the Theory of
Metals} (Benjamin, New York, 1966).

\bibitem{Dong} T. Mayteevarunyoo, B. A. Malomed, and G. Dong, Phys. Rev. A
\textbf{78}, 053601 (2008).

\bibitem{we1} H.~Sakaguchi and B.~A.~Malomed, Phys. Rev. E \textbf{72},
046610 (2005).

\bibitem{NL-1D} F. K. Abdullaev and J. Garnier, Phys. Rev. A \textbf{72},
061605(R) (2005); G. Theocharis, P. Schmelcher, P. G. Kevrekidis, and D. J.
Frantzeskakis, \textit{ibid}. A \textbf{72}, 033614 (2005); Y. Sivan, G.
Fibich, and M. I. Weinstein, Phys. Rev. Lett. \textbf{97}, 193902 (2006); G.
Fibich, Y. Sivan, and M. I. Weinstein, Physica D \textbf{217}, 31 (2006); J.
Belmonte-Beitia, V. M. P\'{e}rez-Garc\'{\i}a, V. Vekslerchik, and P. J. Torres, Phys. Rev.
Lett. \textbf{98}, 064102 (2007); D. A. Zezyulin, G. L. Alfimov, V. V.
Konotop, and V. M. Perez-Garcia, Phys. Rev. A \textbf{76}, 013621 (2007); P.
Niarchou, G. Theocharis, P. G. Kevrekidis, P. Schmelcher, and D. J.
Frantzeskakis, \textit{ibid}. \textbf{76}, 023615 (2007); M. A. Porter, P.
G. Kevrekidis, B. A. Malomed, and D. J. Frantzeskakis, Physica D \textbf{229}%
, 104 (2007); J. Zhou, C. Xue, Y. Qi, and S. Lou, Phys. Lett. A \textbf{372}%
, 4395 (2008); Y. V. Kartashov, V. A. Vysloukh, and L. Torner, Opt. Lett.
\textbf{33}, 1747 (2008); A. S. Rodrigues, P. G. Kevrekidis, M. A. Porter,
D. J. Frantzeskakis, P. Schmelcher, and A. R. Bishop, Phys. Rev. A \textbf{78%
}, 013611 (2008); J. Belmonte-Beitia, V. M. P\'{e}rez-Garc\'{\i}a, V.
Vekslerchik, and P. J. Torres, Discrete Contin. Dyn. Syst. B \textbf{9}, 221
(2008); F. Abdullaev, A. Abdumalikov, and R. Galimzyanov, Phys. Lett. A
\textbf{367}, 149 (2009); V. M. P\'{e}rez-Garc\'{\i}a R. Pardo, Physica D
\textbf{238}, 1352 (2009); F. K. Abdullaev, R. M. Galimzyanov, M. Brtka, and
L. Tomio, Phys. Rev. E \textbf{79}, 056220 (2009).

\bibitem{Kwok} C. H. Tsang, B. Malomed, and K. W. Chow, Discrete Contin.
Dyn. Syst. S, in press.

\bibitem{1delta} D. Witthaut, S. Mossmann, and H. J. Korsch, J. Phys. A:
Math. Gen. \textbf{38}, 1777 (2005).

\bibitem{Azbel} B. A. Malomed and M. Ya. Azbel, Phys. Rev. B \textbf{47},
10402 (1993).

\bibitem{laser} M. I. Rodas-Verde, H. Michinel, and V. M. P\'{e}rez-Garc
\'{\i}a, Phys. Rev. Lett. \textbf{95}, 153903 (2005); A. V. Carpentier, H.
Michinel, M. I. Rodas-Verde, and V. M. P\'{e}rez-Garc\'{\i}a, Phys. Rev. A
\textbf{74}, 013619 (2006).

\bibitem{trap} J. Garnier and F. K. Abdullaev, Phys. Rev. A \textbf{74},
013604 (2006).

\bibitem{NL-1D-2comp} F. K. Abdullaev, A. Gammal, M. Salerno, and L. Tomio,
Phys. Rev. A \textbf{77}, 023615 (2008).

\bibitem{vector} Y. V. Kartashov, B. A. Malomed, V. A. Vysloukh, and L.
Torner, Opt. Lett. \textbf{34}, 3625 (2009).

\bibitem{spatiotemp} J. Belmonte-Beitia, V. M. P\'{e}rez-Garc\'{\i}a, V.
Vekslerchik, and V. V. Konotop, Phys. Rev. Lett. \textbf{100}, 164102 (2008).

\bibitem{NL-2D} H.~Sakaguchi and B.~A.~Malomed, Phys. Rev. E \textbf{72},
046610 (2005); H.~Sakaguchi and B.~A.~Malomed, Phys. Rev. A \textbf{75},
063825 (2007); F. W. Ye, Y. V. Kartashov, B. Hu, and L. Torner, Opt. Exp.
\textbf{17}, 11328 (2009); Y. V. Kartashov, B. A. Malomed, V. A. Vysloukh,
and L. Torner, Opt. Lett. \textbf{34}, 770 (2009).

\bibitem{temporal} L. Berg\'{e}, V. K. Mezentsev, J. Juul Rasmussen, P. L.
Christiansen, and Yu. B. Gaididei, Opt. Lett. \textbf{25}, 1037 (2000); I.
Towers and B. A. Malomed, J. Opt. Soc. Am. \textbf{19}, 537 (2002).

\bibitem{NL-optics} R. Hao, R. Yang, L. Li, and G. Zhou, Opt. Commun.
\textbf{281}, 1256 (2008); Y. V. Kartashov, V. A. Vysloukh, and L. Torner,
Opt. Lett. \textbf{33}, 1774 (2008).

\bibitem{Kominis} Y. Kominis, Phys. Rev. E \textbf{73}, 066619 (2006); Y.
Kominis and K. Hizanidis, Opt. Lett. \textbf{31}, 2888 (2006); Y. Kominis
and K. Hizanidis, Opt. Express \textbf{16}, 12124 (2008).

\bibitem{experiment} Y. V. Kartashov, V. A. Vysloukh, A. Szameit, F.
Dreisow, M. Heinrich, S. Nolte, A. T\"{u}nnermann, T. Pertsch, and L.
Torner, Opt. Lett. \textbf{33}, 1120 (2008).

\bibitem{Panos} Z. Rapti, P. G. Kevrekidis, V. V. Konotop, and C. K. R. T.
Jones, J. Phys. A: Math. Theor. \textbf{40}, 14151 (2007).

\bibitem{Spain} J. Belmonte-Beitia, V. V. Konotop, V. M. P\'{e}rez-Garc\'{\i}%
a. and V. E. Vekslerchik, Chaos, Solitons \& Fractals \textbf{41}, 1158
(2009).

\bibitem{NL-OL2D} Y. Sivan, G. Fibich, B. Ilan, and M. I. Weinstein, Phys.
Rev. E \textbf{78}, 046602 (2008); Y. V. Kartashov, V. A. Vysloukh, and L.
Torner, Opt. Lett. \textbf{33}, 2173 (2008); F. W. Ye, Y. V. Kartashov, B.
Hu, and L. Torner, Opt. Exp. \textbf{17}, 11328 (2009).

\bibitem{PhotCryst} P. Xie, Z.-Q. Zhang, and X. Zhang, Phys. Rev. E \textbf{%
67}, 026607 (2003); A. Ferrando, M. Zacar\'{e}s, P. Fern\'{a}ndez de C\'{o}%
rdoba, D. Binosi, and J. A. Monsoriu, Opt. Express \textbf{11}, 452 (2003);
\textbf{12}, 817 (2004); J. R. Salgueiro, Y. S. Kivshar, D. E. Pelinovsky,
V. Simon, and H. Michinel, Stud. Appl. Math. \textbf{115}, 157 (2005); T.
Mayteevarunyoo and B. A. Malomed, J. Opt. Soc. Am. B \textbf{25}, 1854
(2008); J. R. Salgueiro and Y. S. Kivshar, Eur. Phys. J. Special Topics
\textbf{173}, 281 (2009).

\bibitem{VK} M. Vakhitov and A. Kolokolov, Izvestiya VUZov Radiofizika
\textbf{16}, 1020 (1973) [in Russian; English translation: Radiophys.
Quantum. Electron. \textbf{16}, 783 (1973)]; L. Berg\'{e}, Phys. Rep.
\textbf{303}, 259 (1998).

\bibitem{we2} H.~Sakaguchi and B.~A.~Malomed, J. Phys. B: At. Mol. Opt.
Phys. \textbf{37}, 1443 (2004).

\bibitem{Arik} A. Gubeskys, B. A. Malomed, and I. M. Merhasin, Stud. Appl.
Math. \textbf{115}, 255 (2005).

\bibitem{Kivshar} P. J. Y. Louis, E. A. Ostrovskaya, C. M. Savage, and Y. S.
Kivshar, Phys. Rev. A \textbf{67}, 013602 (2003).

\bibitem{we3} H. Sakaguchi and B. A. Malomed, Phys. Rev. E \textbf{74},
026601 (2006).

\bibitem{Salerno} F. K. Abdullaev and M. Salerno, Phys. Rev. A \textbf{72},
033617 (2005).
\end{thebibliography}
\end{document}